\documentclass[sigconf]{acmart}
\usepackage{graphicx}
\usepackage{subcaption}

\AtBeginDocument{%
  }

\setcopyright{acmlicensed}
\copyrightyear{2024}
\acmYear{2024}
\acmDOI{XXXXXXX.XXXXXXX}

\acmConference[Group '25]{The 2025 ACM International Conference on Supporting Group Work}{January 12--15,
  2025}{Hilton Head Island, South Carolina, USA}
\acmISBN{978-1-4503-XXXX-X/18/06}

\begin{document}

\title[NaviGPT: A Real-Time AI-Driven Mobile Navigation System for PVI]{Enhancing the Travel Experience for People with Visual Impairments through Multimodal Interaction: NaviGPT, A Real-Time AI-Driven Mobile Navigation System}

\author{He Zhang}
\email{hpz5211@psu.edu}
\orcid{0000-0002-8169-1653}
\affiliation{%
  \institution{College of Information Sciences and Technology, Penn State University}
  \city{University Park}
  \state{Pennsylvania}
  \country{USA}
  \postcode{16802}
}
\author{Nicholas J. Falletta}
\orcid{0009-0005-9330-9692}
\email{njf5380@psu.edu}
\affiliation{%
  \institution{College of Information Sciences and Technology, Penn State University}
  \city{University Park}
  \state{Pennsylvania}
  \country{USA}
  \postcode{16802}
}
\author{Jingyi Xie}
\email{jzx5099@psu.edu}
\orcid{0000-0002-2753-2360}
\affiliation{%
  \institution{College of Information Sciences and Technology, Penn State University}
  \city{University Park}
  \state{Pennsylvania}
  \country{USA}
  \postcode{16802}
}
\author{Rui Yu}
\email{rui.yu@louisville.edu}
\orcid{0000-0002-0946-6769}
\affiliation{%
 \institution{Department of Computer Science and Engineering, University of Louisville}
  \city{Louisville}
  \state{KY}
  \country{USA}
  \postcode{40292}
}
\author{Sooyeon Lee}
\email{sooyeon.lee@njit.edu}
\orcid{0000-0002-4971-2004}
\affiliation{
    \institution{Ying Wu College of Computing, New Jersey Institute of Technology}
  \city{Newark}
  \state{NJ}
  \country{USA}
  \postcode{07102}
}
\author{Syed Masum Billah}
\email{sbillah@psu.edu}
\orcid{0000-0001-5063-3808}
\affiliation{%
  \institution{College of Information Sciences and Technology, Penn State University}
  \city{University Park}
  \state{Pennsylvania}
  \country{USA}
  \postcode{16802}
}
\author{John M. Carroll}
\authornote{Corresponding author.}
\orcid{0000-0001-5189-337X}
\email{jmc56@psu.edu}
\affiliation{%
  \institution{College of Information Sciences and Technology, Penn State University}
  \city{University Park}
  \state{Pennsylvania}
  \country{USA}
  \postcode{16802}
}

\renewcommand{\shortauthors}{Zhang et al.}

\begin{abstract}
  Assistive technologies for people with visual impairments (PVI) have made significant advancements, particularly with the integration of artificial intelligence (AI) and real-time sensor technologies. However, current solutions often require PVI to switch between multiple apps and tools for tasks like image recognition, navigation, and obstacle detection, which can hinder a seamless and efficient user experience. In this paper, we present NaviGPT, a high-fidelity prototype that integrates LiDAR-based obstacle detection, vibration feedback, and large language model (LLM) responses to provide a comprehensive and real-time navigation aid for PVI. Unlike existing applications such as Be My AI and Seeing AI, NaviGPT combines image recognition and contextual navigation guidance into a single system, offering continuous feedback on the user's surroundings without the need for app-switching. Meanwhile, NaviGPT compensates for the response delays of LLM by using location and sensor data, aiming to provide practical and efficient navigation support for PVI in dynamic environments.
\end{abstract}

\ccsdesc[500]{Human-centered computing~Ubiquitous and mobile computing systems and tools}
\ccsdesc[500]{Human-centered computing~Accessibility systems and tools}
\ccsdesc[300]{Human-centered computing~Contextual design}

\keywords{People with visual impairments, prototype, AI-assisted tool, accessibility, llm, disability, navigation, multimodal interaction, mobile application}


\begin{teaserfigure}
\centering
\includegraphics[width=0.9\textwidth]{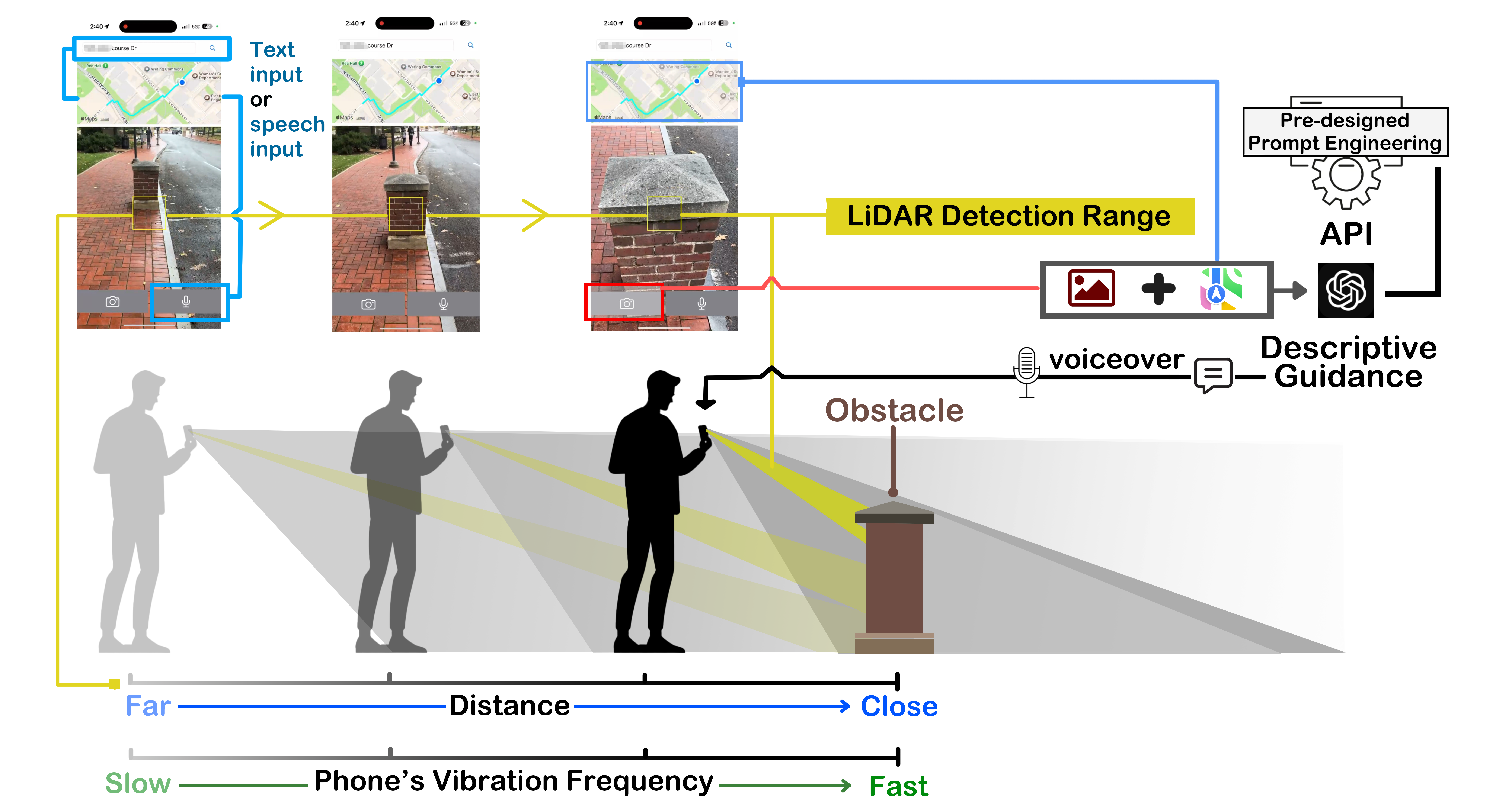}
\caption{Workflow of NaviGPT. }
\Description{The image illustrates the workflow of the NaviGPT system designed to assist people with visual impairments (PVI) in navigating their surroundings through a combination of LiDAR, vibration feedback, and AI-generated guidance. At the top, two mobile screens show the interface where users can input a destination using either text or speech. This activates the navigation system, which displays a walking route on a map. The central part of the image focuses on LiDAR detection, depicted as a yellow detection zone scanning the path in front of the user. The LiDAR detection identifies obstacles, shown as a post in the user's path, marked with a yellow square on the phone screen. Below, a PVI user holds the phone and receives real-time feedback through vibration. The vibration frequency increases as the distance between the user and the obstacle decreases. This is represented on a spectrum, with blue indicating a far distance (slow vibration) and green indicating proximity (fast vibration). On the right, a pre-designed prompt engineering pipeline connects to an API (GPT-4), which processes the obstacle data and provides descriptive guidance through voiceover. This guidance informs the user of the obstacle and offers navigational suggestions, ensuring safe passage. The image emphasizes how NaviGPT integrates LiDAR, tactile feedback, and AI-generated responses to provide a real-time, user-friendly navigation experience for PVI.}
\label{fig.workflow}
\end{teaserfigure}
\maketitle

\section{Introduction}
People with visual impairments (PVI) face challenges in daily life, especially when traveling~\cite{10041898}. They may find themselves in difficult situations due to their inability to effectively grasp their surroundings and changes within them, particularly when in unfamiliar environments, which increases the risk~\cite{10.1145/3315002.3317561}. Even with some assistive tools, such as guide dogs, white canes, tactile paving, and human assistance, a range of issues may still arise, including limitations in usability, range, and interactivity~\cite{10.1145/2470654.2481290}. To reduce the limitations of a single assistive method, it is common for PVI to use multiple assistive tools simultaneously~\cite{kuriakose2022tools}. This combination of assistive technologies can effectively enhance their perception of the environment and improve their daily life experiences, such as the use of both a white cane and tactile paving~\cite{vsakaja2020non}, or a guide dog in conjunction with a white cane~\cite{wiggett2008experience}. Additionally, multimodal and perception-enhancing methods, such as screen magnifiers and screen readers~\cite{leporini2004increasing}, can effectively improve their perceptual efficacy. In this context, developing more usable assistive tools with multiple capabilities for PVI is essential.

With the development of technologies such as computer vision (CV)~\cite{voulodimos2018deep} and natural language processing (NLP)~\cite{khurana2023natural}, applications that replace human vision by using electronic cameras or digital conversions to obtain information are becoming increasingly prevalent. These include functions like object recognition, person detection, text extraction, text reading, and voice assistants. In recent years, with the rise of general artificial intelligence and large language models, these models, such as GPT, are being deployed in more practical applications, bringing benefits to PVI. These models allow PVI to access more content through various interactive means. Applications like ChatGPT~\footnote{\url{https://openai.com/chatgpt/}}, Seeing AI~\footnote{\url{https://www.seeingai.com/}}, Envision AI~\footnote{\url{https://www.letsenvision.com/}}, and Be My AI~\footnote{\url{https://www.bemyeyes.com/blog/introducing-be-my-ai}} utilize a combination of image, text, and voice interactions to provide PVI with descriptions of real-world scenes. This efficient, accurate, and detail-rich descriptive capability, combined with user-friendly natural language interactions, offers tremendous support for accessibility for PVI~\cite{xie2024emergingpracticeslargemultimodal}.



However, although these applications have made significant contributions to the development of assistive tools for PVI, at present, they still lack design considerations specifically tailored for PVI in terms of functionality and interaction methods, or their interactive capabilities are confined within the scope of these applications, lacking interaction with the external environment and various other needs. Moreover, these applications have insufficient support for certain specific scenarios, such as navigation during travel for PVI.

Taking ChatGPT, one of the most popular LLM applications today, and Be My AI, an application specifically designed for visually impaired users, as examples, researchers observed the challenges these applications face in such scenarios~\cite{xie2024emergingpracticeslargemultimodal}. In the case of ChatGPT, it primarily operates via text-based dialogue (prompt engineering). Although it allows interaction through multimodal data like uploaded images, users often need to manually write prompts, which can be complex and time-consuming~\cite{zhang2024redefiningqualitativeanalysisai}, in particular for PVI~\cite{10.1145/3046785}. Even though ChatGPT has introduced voice interaction capabilities, this is limited to a single modality and does not effectively support the navigation needs of PVI. 

For the current version of Be My AI (as of Aug, 2024), it is designed to better meet the needs of PVI (through camera and voice interaction) and enhances their independence. However, past researches have reported some shortcomings in its use, particularly issues such as delays, lengthy feedback, and the requirement for users to actively initiate interactions [reference], which are considered disadvantages in navigation scenarios for PVI. 

To address these limitations and improve the navigation experience of PVI during daily travel, we integrated the LLM (GPT-4) with Apple Maps and developed a high-fidelity prototype called NaviGPT, providing a novel approach to enhancing the PVI experience. By utilizing real-time map navigation information and offering contextual feedback through location data, the LLM delivers a more dynamic and context-aware experience. Unlike interactions initiated by the user for a specific purpose, this system offers an "introduction" feature based on location, allowing PVI to interact with their surroundings in a smoother and more natural way. In this system, Apple Maps provides key information such as nearby landmarks, routes, and the user’s current position, while the LLM interprets these crucial details and offers visual information through the camera during navigation. This enhances both the independence and safety of PVI during travel.

\section{Related Work}

\subsection{The Current State of Assistive Technologies for PVI}

The well-being of PVI has been a major focus for researchers, leading to the continuous development of assistive technologies. Traditional aids such as white canes, guide dogs, and tactile paving have long been used to assist PVI in their life. Currently, updated versions of traditional assistive tools that integrate technology are emerging one after another, such as PVI assistance robots~\cite{10.1145/3555582} and smart white canes~\cite{khan2018technology,10.1145/1639642.1639693}. In addition, with the advent of the internet, mobile technology and wearable device, more advanced solutions have emerged, such as remote volunteer services~\cite{10.1145/3134753,10.1145/3563657.3596019}, where sighted volunteers provide real-time assistance, and specialized assisting systems~\cite{10.1145/3432196} designed to enhance mobility and understanding for PVI. In recent years, the integration of CV and AI has significantly advanced assistive technologies. An increasing number of smart assistive tools are being developed, and in addition, researchers are paying more attention to the collaboration between PVI and these tools and systems, as well as the user experience~\cite{10.1145/3613904.3642030}. In this paper, we mainly focus on navigation tasks for PVI.

\subsection{Navigation Systems for the People with Visual Impairments}

Researchers have developed various prototypes to support PVI in navigating both outdoors and indoors~\cite{survey_of_navigation_aid_2019}. These aids typically include two crucial features for independent mobility: obstacle avoidance and wayfinding~\cite{rafian2017remote}. Obstacle avoidance ensures that PVI can safely navigate their environment without encountering obstacles, often using traditional methods such as guide dogs and white canes. Wayfinding, in contrast, helps PVI to identify and follow a path to a specific location, requiring an understanding of their surroundings through digital or cognitive maps~\cite{cognitive_colalge_1993}, and accurate localization to track their movement within these maps.

The advent of smartphone-based applications like Google Maps~\cite{google_map_app}, BlindSquare~\cite{BlindSquare2020}, and others has significantly enhanced outdoor navigation using Global Positioning System (GPS) and mapping services such as the Google Maps Platform~\cite{google_map_app_api} and OpenStreet Map~\cite{openstreet_map_api}. However, the accuracy of GPS can falter by up to ±5 meters~\cite{GPS2020}, which poses challenges, especially in ``last-few-meters'' navigation~\cite{saha2019closing}. Indoor environments exacerbate these challenges due to poor GPS reception and the absence of detailed indoor mapping~\cite{rodrigo2009robust, li2010indoor}.

To address these challenges, researchers have suggested integrating GPS with other smartphone built-in sensors like Bluetooth~\cite{sato2017navcog3}and Near-field communication (NFC)~\cite{ganz2014percept}, and creating rich indoor maps to capture environmental semantics~\cite{Elmannai2017SensorBasedAD}. Despite the potential of these technologies, they require significant initial investment and ongoing maintenance ~\cite{fallah2012user,bai2014landmark, perez2017assessment} to be effective and also depend on users carrying additional devices such as IR tag readers~\cite{legge2013indoor}.

In recent advancements, the application of CV technologies has emerged as a cost-effective approach for enhancing indoor navigation~\cite{budrionis2020smartphone,yu2024human}. Using smartphones, CV-based systems can interpret visual cues like object recognition~\cite{Zientara2017thirdEye}, color codes, and significant landmarks or signage~\cite{ saha2019closing, fusco2020indoor}. These systems can also process various tags like barcodes, RFID, or vanishing points for better navigation support~\cite{McDaniel2008RFID, Tekin2010Barcode, elloumi2013indoor}. 
However, the reliability of solely using CV for precise navigation for PVI remains insufficient~\cite{saha2019closing}. 
Our prototype integrates CV, particularly utilizing LiDAR, with a LLM to jointly provide assistance for PVI during their travels.


\section{Prototype Design and Implementation}
\subsection{System Architecture}
NaviGPT's architecture is designed to seamlessly integrate various components, creating a robust and responsive navigation system for visually impaired users. The system consists of 7 primary modules:
(1) \textbf{User Interface (UI) Layer}: A simplified, accessible interface optimized for voice and text inputs. (2) \textbf{Navigation Engine}: Built on the Apple Maps API for accurate location and routing services. (3) \textbf{LiDAR Module}: Utilize the LiDAR sensor of the device (iPhone 12 Pro and advanced models) to detect the distance information between the camera and the object. (4) \textbf{Vibration Module}: Invoke the device’s vibration module and use different vibration frequencies to inform the user about the proximity between the device and the object. The closer the object, the higher the vibration frequency; the farther the object, the lower the vibration frequency. (5) \textbf{Image Capture Module}: Utilizes the device's camera to capture environmental images. (6) \textbf{LLM Integration}: Incorporates OpenAI's GPT-4 for image description, information processing, and content generation. (7) \textbf{Data Integration Layer}: Combines inputs from various sources to provide comprehensive navigation assistance.

\subsection{User Interface Design}
The UI of NaviGPT is designed with simplicity and intuitiveness in mind, catering specifically to the needs of visually impaired users. The interface consists of four main components strategically positioned for easy access (as shown in Figure 3): \textbf{Map Interface}: Located at the top of the screen, this simplified map view provides a visual reference for sighted assistants or users with partial vision. \textbf{Voice Interaction Button}: Positioned in the bottom right of the screen, this prominent microphone button allows users to easily activate voice commands and receive audio feedback. \textbf{Camera Interaction Button}: Placed at the bottom left of the screen, this button enables users to quickly capture images of their surroundings for AI analysis. \textbf{LiDAR Detection}: The LiDAR detection feature activates along with the always-on camera, with the detection area being the yellow square at the center. It can measure the distance between the camera and the object selected within the yellow square.

This minimalist design approach ensures that visually impaired users can interact with the system efficiently through touch and voice, reducing cognitive load and enhancing usability.


\subsection{LLM Integration with Navigation and Image Data}
The core of NaviGPT's intelligent navigation system lies in its unique integration of OpenAI's GPT-4 with navigation (from Apple Maps) and image data (from user's input):

\textbf{Map Data Processing}: Apple Maps API provides real-time location data, routing information, and points of interest (destination).

\textbf{Context Generation}: The system combines map data with photos taken by the camera and submits them to GPT-4 via an API. The map provides rich information for navigation tasks, such as the current location of BVI, the navigation route, and the destination, while the photos provide information about the current environment. This information offers the LLM extensive contextual details, enabling it to provide feedback based on the user's purpose and current situation.
\textbf{Intelligent Response Generation}: GPT-4 processes the query and generates natural language responses, providing navigation instructions, environmental descriptions, and safety alerts. Since each time of user input is dynamic, GPT-4 does not provide highly formatted feedback like that of traditional navigation systems (e.g., ``In 200 feet, turn left''). However, we have used prompt engineering method to enhance the formatting of the LLM's responses. This helps the LLM focus more effectively on handling tasks within the navigation context, allowing it to deliver targeted and efficient feedback.

\textbf{Response Refinement}: The system filters and refines the LLM's output to ensure relevance and accuracy before presenting it to the user. As we mentioned before, we use preset prompt engineering methods to control the output. Unlike some LLM-powered applications such as Be My AI or ChatGPT, NaviGPT does not provide detailed descriptions of the images provided by BVI. This avoids lengthy responses, making navigation, an inherently dynamic task that requires some degree of real-time feedback, more concise and efficient.

\subsection{Technical Implementation and Interaction Flow}
NaviGPT's workflow (see in Fig.~\ref{fig.workflow}) is designed to provide a seamless and intuitive experience:

\begin{figure*}[t!]
\centering
\includegraphics[width=0.9\textwidth]{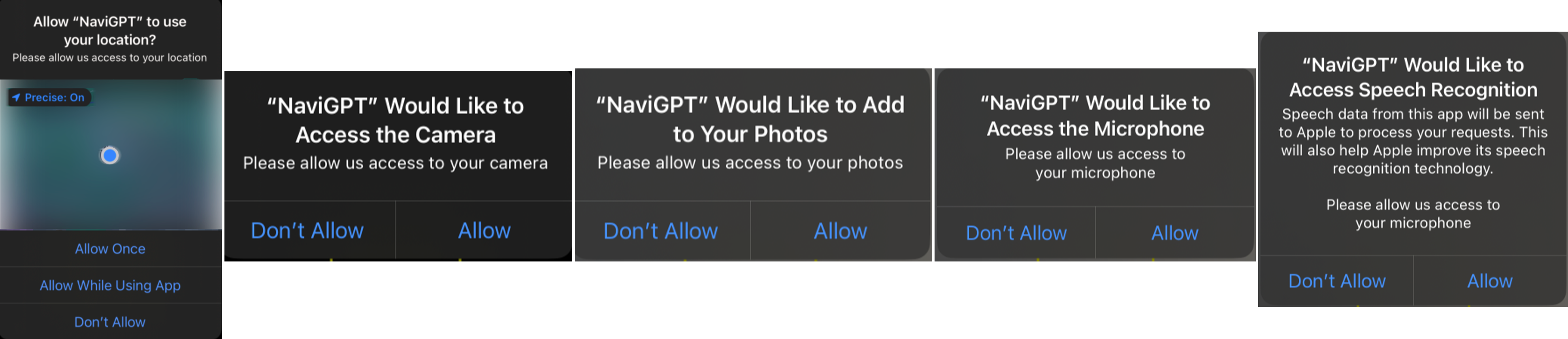}
\caption{NaviGPT requests permissions.}
\label{fig.permissions}
\end{figure*}

\textbf{Initialization}: Upon launch, the application will request the necessary permissions (see in Fig.~\ref{fig.permissions}), including GPS location (for map navigation), camera (for photo interactions), photo library (for storing travel photos), microphone (for voice interactions), and speech recognition (for converting speech to text). It is important to note that these permissions are required to fulfill the app's interaction and functionality needs, and the developers will not access this data in any way. 

\textbf{Destination Input}: Users can input their destination via voice command or text. The speech is converted to text and processed by Apple Maps to extract the address. Apple Maps API is queried to validate the address and create a walking route. This interaction is similar to the process when the user individually uses Apple Maps. After this step, the user will continuously receive general navigation feedback from Apple Maps based on their current location. This feedback will include turn-by-turn directions, real-time updates on their route progress, and alerts for upcoming turns or changes in the path. The navigation will adjust dynamically as the user moves, ensuring they stay on the correct route, with the system providing appropriate guidance for reaching the destination.

\textbf{Real-time LiDAR Detection and Dynamic Vibration Frequency’s Feedback}: The LiDAR will remain active after NaviGPT starts, continuously and in real-time detecting the distance between the objects in the camera's field of view (within the central yellow square) and the mobile device. The device will provide ongoing vibration feedback based on the proximity of the object. A dynamic vibration frequency curve is set according to the distance, with a usable threshold. Specifically, when an object is detected at 10 meters or more, the device will vibrate at the slowest frequency (once every 3 seconds). When an object is detected at 30 cm or less, the device will vibrate at the fastest frequency (5 times per second). Between 10 meters and 30 cm, the vibration frequency decreases as the distance increases, and increases as the distance decreases.

\textbf{Environmental Data Capture}: The user captures their surroundings using the camera button. In this step, when the user wants to explore their surroundings or notices an approaching obstacle through LiDAR detection and vibration feedback, they can quickly take a picture using the camera button in the UI. The photo will be automatically saved to their device's photo album, and further processing will be carried out via an API call to GPT-4 for enhanced insights or contextual understanding.

\textbf{Context Integration}: The system aggregates: a) The captured image. b) Current location (from Apple Maps). c) Destination and next navigation step (from route planning of Apple Maps). This integrated data is combined and then transmitted to GPT-4 all at once.

\begin{figure}[h]
    \centering
    \begin{subfigure}{0.3\textwidth}
    \includegraphics[width=\textwidth]{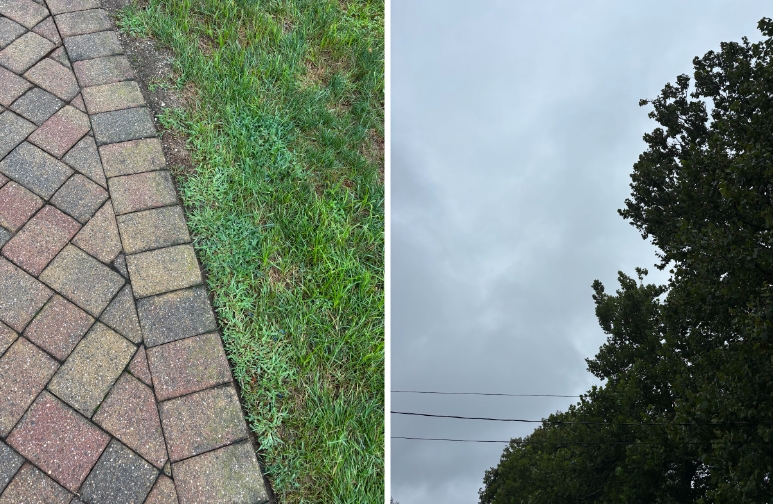}
    \caption{Example of Bad Photos.}
    \label{fig.badphotos}
     \end{subfigure}
     \hfill
    \begin{subfigure}{0.3\textwidth}
    \includegraphics[width=\textwidth]{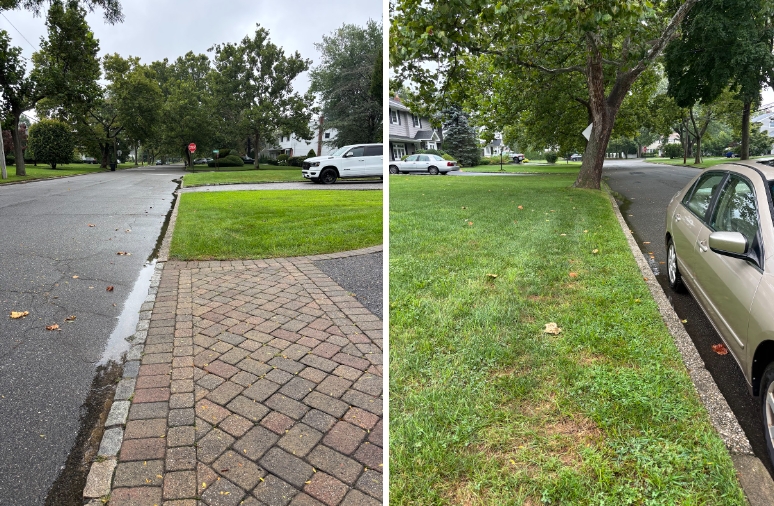}
    \caption{Example of Good Photos.}
    \label{fig.goodphotos1}
    \end{subfigure}
    \hfill
    \begin{subfigure}{0.45\textwidth}
    \includegraphics[width=\textwidth]{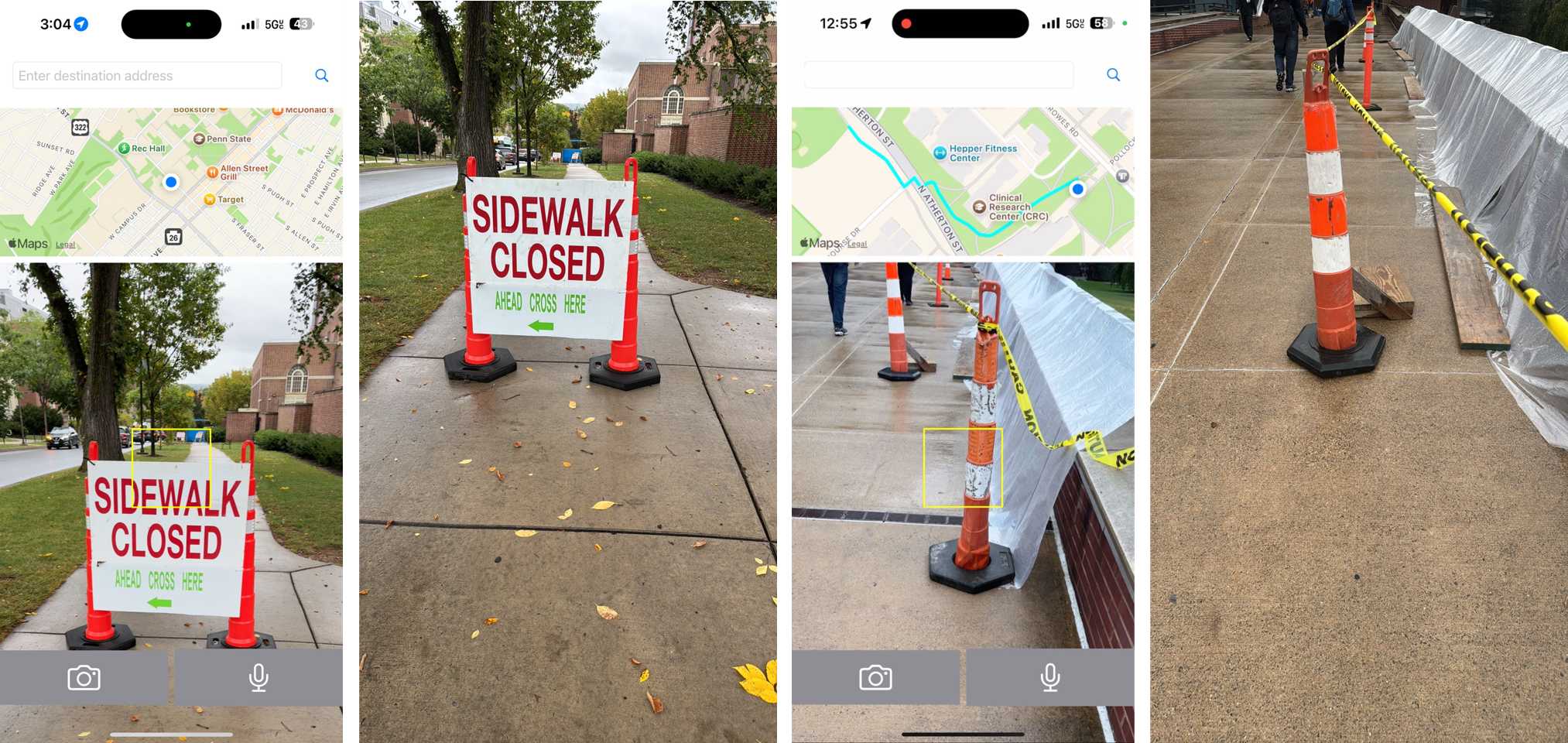}
    \caption{Example of UI and Good Photos.}
    \label{fig.goodphotos2}
    \end{subfigure}
    \caption{Example of Bad and Good Photos}
    \end{figure}

\textbf{LLM Processing and Response Generation}: GPT-4 processes the multimodal prompt (examples are shown in Fig.~\ref{fig.goodphotos1}, and~\ref{fig.goodphotos2}). It generates a natural language response that includes: (a) Description of the captured image. (b) Current location description. (c) Safety assessment of the immediate environment. (d) Next navigation instruction. (e) Any relevant warnings or additional information. If the captured photo has a poor angle or does not detect content related to navigation (such as in Fig.~\ref{fig.badphotos}), NaviGPT will still provide a response to the user's input but will additionally prompt the user to retake a photo that is more suitable for navigation purposes. Notably, in this scenario, the GPT feedback audio will take priority over the general navigation prompts. Once the GPT feedback has finished playing, the regular navigation instructions will resume. Importantly, the navigation functionality will not be interrupted during the photo-taking process or while receiving the GPT feedback.
    

\textbf{Continuous Monitoring and Updates}: The system continuously updates the user's location. It prompts for new image captures at key decision points or at regular intervals. The process repeats from step 3 to provide ongoing navigation assistance.




\section{Advantages of NaviGPT Over Existing Systems}
\subsection{Comparison with Existing Map Navigation Systems.}
NaviGPT, developed based on Apple Maps, goes beyond the standard features of mainstream navigation systems like Google Maps and Apple Maps. While these platforms offer extensive functions, their complexity often overwhelms PVI. Designed primarily for sighted users, these tools require navigating intricate menus and visual prompts, which can hinder PVI users.

NaviGPT, however, is tailored specifically to PVI needs, prioritizing simplicity and accessibility. Instead of relying on visual elements, detailed maps, and text-based instructions, it combines LiDAR, vibration feedback, and contextual responses from a LLM, reducing the dependency on vision-based interactions. This allows PVI users to focus on navigation without processing complex visual data.

A key advantage of NaviGPT is its simplified UI, offering fast, spoken instructions with minimal distractions. Unlike existing apps that require multiple actions to configure settings or navigate cluttered screens, NaviGPT delivers clear and timely feedback. By providing focused guidance rather than excessive detail, it ensures a smoother, more efficient navigation experience for PVI users, making it easier to access essential information and move safely in dynamic environments.

\subsection{Comparison with Existing AI-powered Assistive Systems for PVI.}
By comparing NaviGPT with other LLM-integrated applications like Be My AI and Seeing AI, NaviGPT’s primary advantage lies in its seamless integration of a navigation system into the LLM interaction flow. It also utilizes LiDAR and vibration feedback to simulate a white cane, providing real-time feedback to PVI. This integration allows users to access key features, AI-based image identification, navigation, and road safety confirmation—without switching between apps, enhancing the overall experience.

Unlike Be My AI, which focuses on detailed image descriptions, NaviGPT is designed for daily travel and navigation, making it more efficient in recognizing and responding to the changing environment. PVI require quick access to safety information, such as detecting obstacles and signage, which NaviGPT provides in real-time. While Be My AI offers more detailed feedback, often exceeding 60 words, it typically requires longer wait and reading out time (over 5 seconds in tests). This level of detail may be unnecessary in dynamic walking scenarios, where rapid feedback is more valuable.

Table~\ref{tab:my-table} compares feedback from Be My AI and NaviGPT. While Be My AI provides more detailed information, NaviGPT’s quicker, focused feedback ensures safe and efficient navigation through LiDAR and vibration alerts.

\begin{table*}[]
\centering
\begin{tabular}{c|cc}
\hline
 &
  \multicolumn{2}{c}{Feedback (Words Count)} \\ \hline
Input Picture &
  \multicolumn{1}{c|}{From Be My AI (86 words)} &
  From NaviGPT (28 words) \\ \hline
 \raisebox{-0.5\height}{\includegraphics[width=0.12\textwidth]{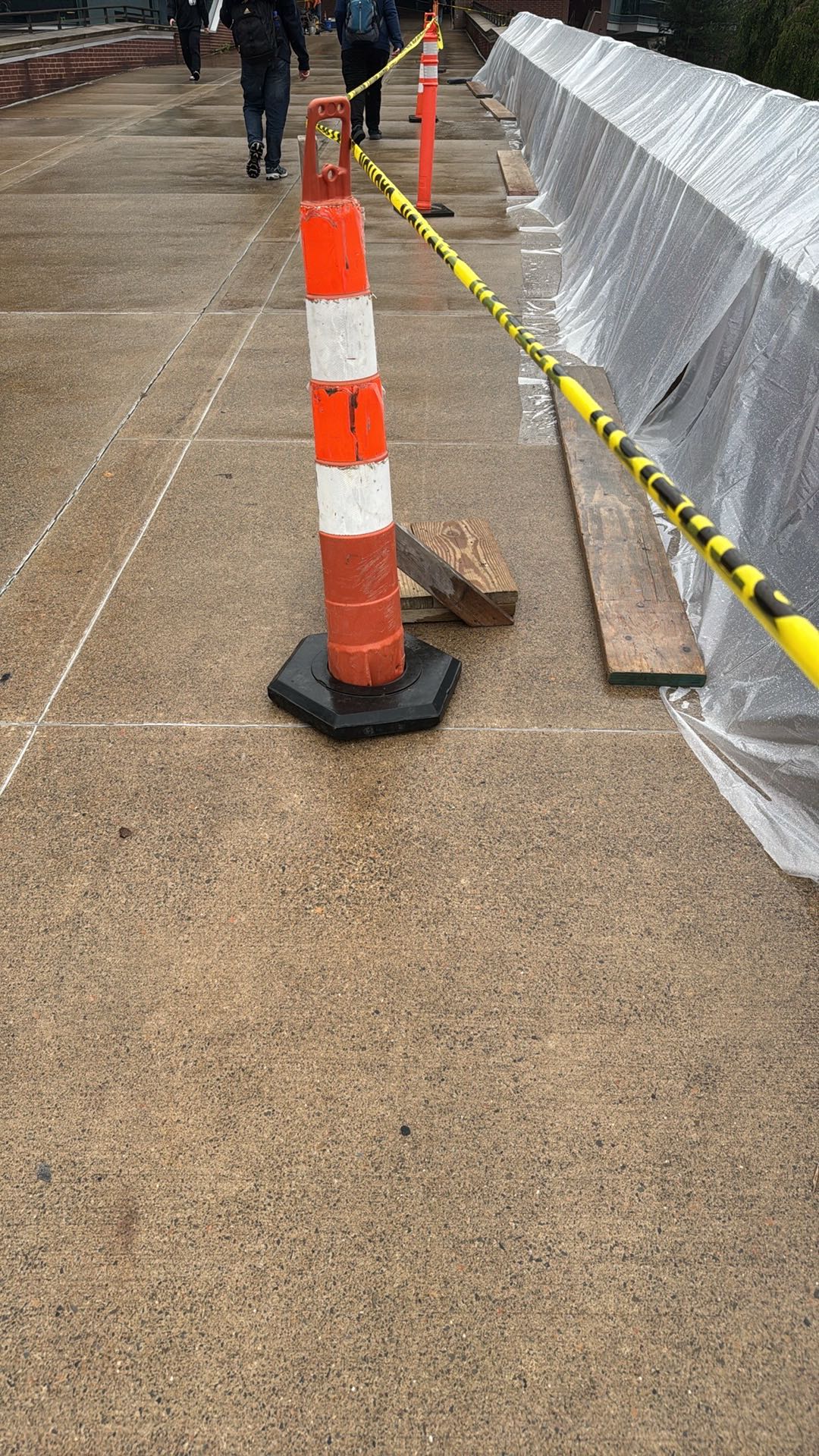}}&
  \multicolumn{1}{c|}{\begin{tabular}[c]{@{}c@{}}The image shows a walkway that appears to be under construction \\ or maintenance. There are orange and white traffic cones \\ placed along the path, with yellow caution tape strung between them,\\  creating a barrier. On the right side, there is a plastic sheet covering \\ something, possibly construction materials or an area under repair. \\ Wooden planks are also visible on the ground near the plastic sheet. \\ In the background, a few people are walking along the path. \\ The ground looks wet, suggesting it might have rained recently.\end{tabular}} &
  \begin{tabular}[c]{@{}c@{}}It's safe to walk \\ but be cautious \\ as there is construction \\ underway with barriers \\ and caution tape \\ indicating a restricted area. \\ Please navigate around the \\ construction area carefully.\end{tabular} \\ \hline
\end{tabular}
\caption{Example of Feedbacks from Be My AI and NaviGPT.}
\label{tab:my-table}
\end{table*}
\raggedbottom

\section{Limitations and Future Work}
While NaviGPT shows promise, there are several limitations that must be addressed in future iterations. One of the main challenges is the system's reliance on external hardware, such as LiDAR sensors, which are only available on certain advanced devices. This restricts accessibility for users who may not have access to the latest technology~\cite{botelho2021accessibility}. Addressing this issue could involve exploring alternative, more widely available sensors or optimizing the system to work without advanced hardware. If we rely solely on AI and reduce dependence on sensors, this could result in delays, especially in environments with poor internet connectivity. Due to the weak GPS signal inside buildings or underground, the current version of the navigation feature is suitable for outdoor environments only. In addition, privacy concerns related to the continuous use of LiDAR and camera inputs must be addressed. Users may feel uncomfortable with persistent data collection, so future versions of NaviGPT will need to ensure strong privacy protections, including local data processing options and transparent data use policies.
Our future work will focus on several key areas to further enhance NaviGPT and fully explore its potential as a navigation aid for PVI, such as adding object recognition and fast feedback features. Additionally, we plan to conduct comprehensive user research to better understand the specific needs and preferences of PVI. These experiments will provide critical insights into how users interact with the application in real-world environments, enabling us to identify potential pain points and improve the system's design, responsiveness, and user experience.

Another important avenue for future development is the integration of more advanced multimodal systems. By further leveraging AI, we aim to bridge the functional gaps between various assistive technologies and consolidate them into a unified platform. This could include expanding the current functionality to incorporate speech-based interactions, real-time environmental mapping, and even predictive analytics that anticipate the user's next actions, movements and emotions based on advanced models and historical data~\cite{10494076}. Such integrations would result in a more intelligent, cohesive system, enhancing the user's ability to navigate complex and dynamic environments.

In addition to improving the technical capabilities, we also plan to explore how the application could adapt to different user contexts, such as indoor navigation in crowded spaces or specialized outdoor environments (e.g., urban vs. rural settings). This would involve customizing feedback based on situational awareness, ensuring the system remains flexible and effective in a wide range of scenarios.
\begin{acks}
This research was supported by the US National Institutes of Health, and the National Library of Medicine (R01 LM013330). 
\end{acks}

\bibliographystyle{ACM-Reference-Format}
\bibliography{main}


\begin{thebibliography}{46}


\ifx \showCODEN    \undefined \def \showCODEN     #1{\unskip}     \fi
\ifx \showDOI      \undefined \def \showDOI       #1{#1}\fi
\ifx \showISBNx    \undefined \def \showISBNx     #1{\unskip}     \fi
\ifx \showISBNxiii \undefined \def \showISBNxiii  #1{\unskip}     \fi
\ifx \showISSN     \undefined \def \showISSN      #1{\unskip}     \fi
\ifx \showLCCN     \undefined \def \showLCCN      #1{\unskip}     \fi
\ifx \shownote     \undefined \def \shownote      #1{#1}          \fi
\ifx \showarticletitle \undefined \def \showarticletitle #1{#1}   \fi
\ifx \showURL      \undefined \def \showURL       {\relax}        \fi
\providecommand\bibfield[2]{#2}
\providecommand\bibinfo[2]{#2}
\providecommand\natexlab[1]{#1}
\providecommand\showeprint[2][]{arXiv:#2}

\bibitem[goo(2021a)]%
        {google_map_app}
 \bibinfo{year}{2021}\natexlab{a}.
\newblock \bibinfo{title}{Google Maps - Transit \& Food}.
\newblock
\newblock
\urldef\tempurl%
\url{https://apps.apple.com/us/app/google-maps-transit-food/id585027354}
\showURL{%
Retrieved February 13, 2021 from \tempurl}


\bibitem[goo(2021b)]%
        {google_map_app_api}
 \bibinfo{year}{2021}\natexlab{b}.
\newblock \bibinfo{title}{Welcome to Google Maps Platform - Explore where real-world insights and immersive location experiences can take your business}.
\newblock
\newblock
\urldef\tempurl%
\url{https://cloud.google.com/maps-platform/}
\showURL{%
Retrieved February 13, 2021 from \tempurl}


\bibitem[ope(2021)]%
        {openstreet_map_api}
 \bibinfo{year}{2021}\natexlab{}.
\newblock \bibinfo{title}{Welcome to OpenStreetMap! OpenStreetMap is a map of the world, created by people like you and free to use under an open license.}
\newblock
\newblock
\urldef\tempurl%
\url{https://www.openstreetmap.org/}
\showURL{%
Retrieved February 13, 2021 from \tempurl}


\bibitem[Ahmetovic et~al\mbox{.}(2019)]%
        {10.1145/3315002.3317561}
\bibfield{author}{\bibinfo{person}{Dragan Ahmetovic}, \bibinfo{person}{Jo\~{a}o Guerreiro}, \bibinfo{person}{Eshed Ohn-Bar}, \bibinfo{person}{Kris~M. Kitani}, {and} \bibinfo{person}{Chieko Asakawa}.} \bibinfo{year}{2019}\natexlab{}.
\newblock \showarticletitle{Impact of Expertise on Interaction Preferences for Navigation Assistance of Visually Impaired Individuals}. In \bibinfo{booktitle}{\emph{Proceedings of the 16th International Web for All Conference}} (San Francisco, CA, USA) \emph{(\bibinfo{series}{W4A '19})}. \bibinfo{publisher}{Association for Computing Machinery}, \bibinfo{address}{New York, NY, USA}, Article \bibinfo{articleno}{31}, \bibinfo{numpages}{9}~pages.
\newblock
\showISBNx{9781450367165}
\urldef\tempurl%
\url{https://doi.org/10.1145/3315002.3317561}
\showDOI{\tempurl}


\bibitem[Bai et~al\mbox{.}(2014)]%
        {bai2014landmark}
\bibfield{author}{\bibinfo{person}{Yicheng Bai}, \bibinfo{person}{Wenyan Jia}, \bibinfo{person}{Hong Zhang}, \bibinfo{person}{Zhi-Hong Mao}, {and} \bibinfo{person}{Mingui Sun}.} \bibinfo{year}{2014}\natexlab{}.
\newblock \showarticletitle{Landmark-based indoor positioning for visually impaired individuals}. In \bibinfo{booktitle}{\emph{2014 12th International Conference on Signal Processing (ICSP)}}. IEEE, \bibinfo{pages}{668--671}.
\newblock


\bibitem[Bhat and Zhao(2022)]%
        {10.1145/3555582}
\bibfield{author}{\bibinfo{person}{Prajna Bhat} {and} \bibinfo{person}{Yuhang Zhao}.} \bibinfo{year}{2022}\natexlab{}.
\newblock \showarticletitle{"I was Confused by It; It was Confused by Me:" Exploring the Experiences of People with Visual Impairments around Mobile Service Robots}.
\newblock \bibinfo{journal}{\emph{Proc. ACM Hum.-Comput. Interact.}} \bibinfo{volume}{6}, \bibinfo{number}{CSCW2}, Article \bibinfo{articleno}{481} (\bibinfo{date}{Nov.} \bibinfo{year}{2022}), \bibinfo{numpages}{26}~pages.
\newblock
\urldef\tempurl%
\url{https://doi.org/10.1145/3555582}
\showDOI{\tempurl}


\bibitem[BlindSquare(2020)]%
        {BlindSquare2020}
\bibfield{author}{\bibinfo{person}{BlindSquare}.} \bibinfo{year}{2020}\natexlab{}.
\newblock \bibinfo{title}{BlindSquare iOS Application}.
\newblock \bibinfo{howpublished}{\url{https://www.blindsquare.com/}}.
\newblock


\bibitem[Boldu et~al\mbox{.}(2020)]%
        {10.1145/3432196}
\bibfield{author}{\bibinfo{person}{Roger Boldu}, \bibinfo{person}{Denys~J.C. Matthies}, \bibinfo{person}{Haimo Zhang}, {and} \bibinfo{person}{Suranga Nanayakkara}.} \bibinfo{year}{2020}\natexlab{}.
\newblock \showarticletitle{AiSee: An Assistive Wearable Device to Support Visually Impaired Grocery Shoppers}.
\newblock \bibinfo{journal}{\emph{Proc. ACM Interact. Mob. Wearable Ubiquitous Technol.}} \bibinfo{volume}{4}, \bibinfo{number}{4}, Article \bibinfo{articleno}{119} (\bibinfo{date}{Dec.} \bibinfo{year}{2020}), \bibinfo{numpages}{25}~pages.
\newblock
\urldef\tempurl%
\url{https://doi.org/10.1145/3432196}
\showDOI{\tempurl}


\bibitem[Botelho(2021)]%
        {botelho2021accessibility}
\bibfield{author}{\bibinfo{person}{Fernando~HF Botelho}.} \bibinfo{year}{2021}\natexlab{}.
\newblock \showarticletitle{Accessibility to digital technology: Virtual barriers, real opportunities}.
\newblock \bibinfo{journal}{\emph{Assistive Technology}} \bibinfo{volume}{33}, \bibinfo{number}{sup1} (\bibinfo{year}{2021}), \bibinfo{pages}{27--34}.
\newblock
\urldef\tempurl%
\url{https://doi.org/10.1080/10400435.2021.1945705}
\showDOI{\tempurl}


\bibitem[Budrionis et~al\mbox{.}(2020)]%
        {budrionis2020smartphone}
\bibfield{author}{\bibinfo{person}{Andrius Budrionis}, \bibinfo{person}{Darius Plikynas}, \bibinfo{person}{Povilas Daniu{\v{s}}is}, {and} \bibinfo{person}{Audrius Indrulionis}.} \bibinfo{year}{2020}\natexlab{}.
\newblock \showarticletitle{Smartphone-based computer vision travelling aids for blind and visually impaired individuals: A systematic review}.
\newblock \bibinfo{journal}{\emph{Assistive Technology}} (\bibinfo{year}{2020}), \bibinfo{pages}{1--17}.
\newblock


\bibitem[El-Taher et~al\mbox{.}(2023)]%
        {10041898}
\bibfield{author}{\bibinfo{person}{Fatma El-Zahraa El-Taher}, \bibinfo{person}{Luis Miralles-Pechuán}, \bibinfo{person}{Jane Courtney}, \bibinfo{person}{Kristina Millar}, \bibinfo{person}{Chantelle Smith}, {and} \bibinfo{person}{Susan Mckeever}.} \bibinfo{year}{2023}\natexlab{}.
\newblock \showarticletitle{A Survey on Outdoor Navigation Applications for People With Visual Impairments}.
\newblock \bibinfo{journal}{\emph{IEEE Access}}  \bibinfo{volume}{11} (\bibinfo{year}{2023}), \bibinfo{pages}{14647--14666}.
\newblock
\urldef\tempurl%
\url{https://doi.org/10.1109/ACCESS.2023.3244073}
\showDOI{\tempurl}


\bibitem[Elloumi et~al\mbox{.}(2013)]%
        {elloumi2013indoor}
\bibfield{author}{\bibinfo{person}{Wael Elloumi}, \bibinfo{person}{Kamel Guissous}, \bibinfo{person}{Aladine Chetouani}, \bibinfo{person}{Rapha{\"e}l Canals}, \bibinfo{person}{R{\'e}my Leconge}, \bibinfo{person}{Bruno Emile}, {and} \bibinfo{person}{Sylvie Treuillet}.} \bibinfo{year}{2013}\natexlab{}.
\newblock \showarticletitle{Indoor navigation assistance with a Smartphone camera based on vanishing points}. In \bibinfo{booktitle}{\emph{International Conference on Indoor Positioning and Indoor Navigation}}. IEEE, \bibinfo{pages}{1--9}.
\newblock


\bibitem[Elmannai and Elleithy(2017)]%
        {Elmannai2017SensorBasedAD}
\bibfield{author}{\bibinfo{person}{Wafa Elmannai} {and} \bibinfo{person}{Khaled~M. Elleithy}.} \bibinfo{year}{2017}\natexlab{}.
\newblock \showarticletitle{Sensor-based assistive devices for visually-impaired people: Current status, challenges, and future directions}.
\newblock \bibinfo{journal}{\emph{Sensors (Basel, Switzerland)}}  \bibinfo{volume}{17} (\bibinfo{year}{2017}).
\newblock


\bibitem[Fallah et~al\mbox{.}(2012)]%
        {fallah2012user}
\bibfield{author}{\bibinfo{person}{Navid Fallah}, \bibinfo{person}{Ilias Apostolopoulos}, \bibinfo{person}{Kostas Bekris}, {and} \bibinfo{person}{Eelke Folmer}.} \bibinfo{year}{2012}\natexlab{}.
\newblock \showarticletitle{The user as a sensor: navigating users with visual impairments in indoor spaces using tactile landmarks}. In \bibinfo{booktitle}{\emph{Proceedings of the SIGCHI Conference on Human Factors in Computing Systems}}. \bibinfo{pages}{425--432}.
\newblock


\bibitem[Fusco and Coughlan(2020)]%
        {fusco2020indoor}
\bibfield{author}{\bibinfo{person}{Giovanni Fusco} {and} \bibinfo{person}{James~M Coughlan}.} \bibinfo{year}{2020}\natexlab{}.
\newblock \showarticletitle{Indoor localization for visually impaired travelers using computer vision on a smartphone}. In \bibinfo{booktitle}{\emph{Proceedings of the 17th International Web for All Conference}}. \bibinfo{pages}{1--11}.
\newblock


\bibitem[Ganz et~al\mbox{.}(2014)]%
        {ganz2014percept}
\bibfield{author}{\bibinfo{person}{Aura Ganz}, \bibinfo{person}{James~M Schafer}, \bibinfo{person}{Yang Tao}, \bibinfo{person}{Carole Wilson}, {and} \bibinfo{person}{Meg Robertson}.} \bibinfo{year}{2014}\natexlab{}.
\newblock \showarticletitle{PERCEPT-II: Smartphone based indoor navigation system for the blind}. In \bibinfo{booktitle}{\emph{2014 36th Annual International Conference of the IEEE Engineering in Medicine and Biology Society}}. IEEE, \bibinfo{pages}{3662--3665}.
\newblock


\bibitem[GPS.gov({[n.\,d.]})]%
        {GPS2020}
\bibfield{author}{\bibinfo{person}{GPS.gov}.} \bibinfo{year}{[n.\,d.]}\natexlab{}.
\newblock \bibinfo{title}{GPS Accuracy}.
\newblock \bibinfo{howpublished}{\url{https://www.gps.gov/systems/gps/performance/accuracy/}}.
\newblock


\bibitem[Ju et~al\mbox{.}(2009)]%
        {10.1145/1639642.1639693}
\bibfield{author}{\bibinfo{person}{Jin~Sun Ju}, \bibinfo{person}{Eunjeong Ko}, {and} \bibinfo{person}{Eun~Yi Kim}.} \bibinfo{year}{2009}\natexlab{}.
\newblock \showarticletitle{EYECane: navigating with camera embedded white cane for visually impaired person}. In \bibinfo{booktitle}{\emph{Proceedings of the 11th International ACM SIGACCESS Conference on Computers and Accessibility}} (Pittsburgh, Pennsylvania, USA) \emph{(\bibinfo{series}{Assets '09})}. \bibinfo{publisher}{Association for Computing Machinery}, \bibinfo{address}{New York, NY, USA}, \bibinfo{pages}{237–238}.
\newblock
\showISBNx{9781605585581}
\urldef\tempurl%
\url{https://doi.org/10.1145/1639642.1639693}
\showDOI{\tempurl}


\bibitem[Khan et~al\mbox{.}(2018)]%
        {khan2018technology}
\bibfield{author}{\bibinfo{person}{Izaz Khan}, \bibinfo{person}{Shah Khusro}, {and} \bibinfo{person}{Irfan Ullah}.} \bibinfo{year}{2018}\natexlab{}.
\newblock \showarticletitle{Technology-assisted white cane: evaluation and future directions}.
\newblock \bibinfo{journal}{\emph{PeerJ}}  \bibinfo{volume}{6} (\bibinfo{year}{2018}), \bibinfo{pages}{e6058}.
\newblock


\bibitem[Khurana et~al\mbox{.}(2023)]%
        {khurana2023natural}
\bibfield{author}{\bibinfo{person}{Diksha Khurana}, \bibinfo{person}{Aditya Koli}, \bibinfo{person}{Kiran Khatter}, {and} \bibinfo{person}{Sukhdev Singh}.} \bibinfo{year}{2023}\natexlab{}.
\newblock \showarticletitle{Natural language processing: state of the art, current trends and challenges}.
\newblock \bibinfo{journal}{\emph{Multimedia tools and applications}} \bibinfo{volume}{82}, \bibinfo{number}{3} (\bibinfo{year}{2023}), \bibinfo{pages}{3713--3744}.
\newblock


\bibitem[Kuriakose et~al\mbox{.}(2022)]%
        {kuriakose2022tools}
\bibfield{author}{\bibinfo{person}{Bineeth Kuriakose}, \bibinfo{person}{Raju Shrestha}, {and} \bibinfo{person}{Frode~Eika Sandnes}.} \bibinfo{year}{2022}\natexlab{}.
\newblock \showarticletitle{Tools and technologies for blind and visually impaired navigation support: a review}.
\newblock \bibinfo{journal}{\emph{IETE Technical Review}} \bibinfo{volume}{39}, \bibinfo{number}{1} (\bibinfo{year}{2022}), \bibinfo{pages}{3--18}.
\newblock
\urldef\tempurl%
\url{https://doi.org/10.1080/02564602.2020.1819893}
\showDOI{\tempurl}


\bibitem[Legge et~al\mbox{.}(2013)]%
        {legge2013indoor}
\bibfield{author}{\bibinfo{person}{Gordon~E Legge}, \bibinfo{person}{Paul~J Beckmann}, \bibinfo{person}{Bosco~S Tjan}, \bibinfo{person}{Gary Havey}, \bibinfo{person}{Kevin Kramer}, \bibinfo{person}{David Rolkosky}, \bibinfo{person}{Rachel Gage}, \bibinfo{person}{Muzi Chen}, \bibinfo{person}{Sravan Puchakayala}, {and} \bibinfo{person}{Aravindhan Rangarajan}.} \bibinfo{year}{2013}\natexlab{}.
\newblock \showarticletitle{Indoor navigation by people with visual impairment using a digital sign system}.
\newblock \bibinfo{journal}{\emph{PloS one}} \bibinfo{volume}{8}, \bibinfo{number}{10} (\bibinfo{year}{2013}).
\newblock


\bibitem[Leporini and Patern{\`o}(2004)]%
        {leporini2004increasing}
\bibfield{author}{\bibinfo{person}{Barbara Leporini} {and} \bibinfo{person}{Fabio Patern{\`o}}.} \bibinfo{year}{2004}\natexlab{}.
\newblock \showarticletitle{Increasing usability when interacting through screen readers}.
\newblock \bibinfo{journal}{\emph{Universal access in the information society}}  \bibinfo{volume}{3} (\bibinfo{year}{2004}), \bibinfo{pages}{57--70}.
\newblock


\bibitem[Li and Lee(2010)]%
        {li2010indoor}
\bibfield{author}{\bibinfo{person}{Ki-Joune Li} {and} \bibinfo{person}{Jiyeong Lee}.} \bibinfo{year}{2010}\natexlab{}.
\newblock \showarticletitle{Indoor spatial awareness initiative and standard for indoor spatial data}. In \bibinfo{booktitle}{\emph{Proceedings of IROS 2010 Workshop on Standardization for Service Robot}}, Vol.~\bibinfo{volume}{18}.
\newblock


\bibitem[McDaniel et~al\mbox{.}(2008)]%
        {McDaniel2008RFID}
\bibfield{author}{\bibinfo{person}{Troy McDaniel}, \bibinfo{person}{Kanav Kahol}, \bibinfo{person}{Daniel Villanueva}, {and} \bibinfo{person}{Sethuraman Panchanathan}.} \bibinfo{year}{2008}\natexlab{}.
\newblock \showarticletitle{Integration of RFID and computer vision for remote object perception for individuals who are blind}. In \bibinfo{booktitle}{\emph{Proceedings of the 2008 Ambi-Sys Workshop on Haptic User Interfaces in Ambient Media Systems, HAS 2008}}. \bibinfo{publisher}{Association for Computing Machinery, Inc}.
\newblock
\newblock
\shownote{2008 1st Ambi-Sys Workshop on Haptic User Interfaces in Ambient Media Systems, HAS 2008 ; Conference date: 11-02-2008 Through 14-02-2008}.


\bibitem[Nicolau et~al\mbox{.}(2017)]%
        {10.1145/3046785}
\bibfield{author}{\bibinfo{person}{Hugo Nicolau}, \bibinfo{person}{Kyle Montague}, \bibinfo{person}{Tiago Guerreiro}, \bibinfo{person}{Andr\'{e} Rodrigues}, {and} \bibinfo{person}{Vicki~L. Hanson}.} \bibinfo{year}{2017}\natexlab{}.
\newblock \showarticletitle{Investigating Laboratory and Everyday Typing Performance of Blind Users}.
\newblock \bibinfo{journal}{\emph{ACM Trans. Access. Comput.}} \bibinfo{volume}{10}, \bibinfo{number}{1}, Article \bibinfo{articleno}{4} (\bibinfo{date}{March} \bibinfo{year}{2017}), \bibinfo{numpages}{26}~pages.
\newblock
\showISSN{1936-7228}
\urldef\tempurl%
\url{https://doi.org/10.1145/3046785}
\showDOI{\tempurl}


\bibitem[Pan\"{e}els et~al\mbox{.}(2013)]%
        {10.1145/2470654.2481290}
\bibfield{author}{\bibinfo{person}{Sabrina~A. Pan\"{e}els}, \bibinfo{person}{Adriana Olmos}, \bibinfo{person}{Jeffrey~R. Blum}, {and} \bibinfo{person}{Jeremy~R. Cooperstock}.} \bibinfo{year}{2013}\natexlab{}.
\newblock \showarticletitle{Listen to it yourself! evaluating usability of what's around me? for the blind}. In \bibinfo{booktitle}{\emph{Proceedings of the SIGCHI Conference on Human Factors in Computing Systems}} (Paris, France) \emph{(\bibinfo{series}{CHI '13})}. \bibinfo{publisher}{Association for Computing Machinery}, \bibinfo{address}{New York, NY, USA}, \bibinfo{pages}{2107–2116}.
\newblock
\showISBNx{9781450318990}
\urldef\tempurl%
\url{https://doi.org/10.1145/2470654.2481290}
\showDOI{\tempurl}


\bibitem[P{\'e}rez et~al\mbox{.}(2017)]%
        {perez2017assessment}
\bibfield{author}{\bibinfo{person}{J~Eduardo P{\'e}rez}, \bibinfo{person}{Myriam Arrue}, \bibinfo{person}{Masatomo Kobayashi}, \bibinfo{person}{Hironobu Takagi}, {and} \bibinfo{person}{Chieko Asakawa}.} \bibinfo{year}{2017}\natexlab{}.
\newblock \showarticletitle{Assessment of semantic taxonomies for blind indoor navigation based on a shopping center use case}. In \bibinfo{booktitle}{\emph{Proceedings of the 14th Web for All Conference on The Future of Accessible Work}}. \bibinfo{pages}{1--4}.
\newblock


\bibitem[Rafian and Legge(2017)]%
        {rafian2017remote}
\bibfield{author}{\bibinfo{person}{Paymon Rafian} {and} \bibinfo{person}{Gordon~E Legge}.} \bibinfo{year}{2017}\natexlab{}.
\newblock \showarticletitle{Remote sighted assistants for indoor location sensing of visually impaired pedestrians}.
\newblock \bibinfo{journal}{\emph{ACM Transactions on Applied Perception (TAP)}} \bibinfo{volume}{14}, \bibinfo{number}{3} (\bibinfo{year}{2017}), \bibinfo{pages}{19}.
\newblock


\bibitem[Real and Araujo(2019)]%
        {survey_of_navigation_aid_2019}
\bibfield{author}{\bibinfo{person}{Santiago Real} {and} \bibinfo{person}{Alvaro Araujo}.} \bibinfo{year}{2019}\natexlab{}.
\newblock \showarticletitle{Navigation systems for the blind and visually impaired: Past work, challenges, and open problems}.
\newblock \bibinfo{journal}{\emph{Sensors (Basel, Switzerland)}} \bibinfo{volume}{19}, \bibinfo{number}{15} (\bibinfo{date}{02 Aug} \bibinfo{year}{2019}), \bibinfo{pages}{3404}.
\newblock
\showISSN{1424-8220}
\urldef\tempurl%
\url{https://doi.org/10.3390/s19153404}
\showDOI{\tempurl}
\newblock
\shownote{31382536[pmid]}.


\bibitem[Rodrigo et~al\mbox{.}(2009)]%
        {rodrigo2009robust}
\bibfield{author}{\bibinfo{person}{Ranga Rodrigo}, \bibinfo{person}{Mehrnaz Zouqi}, \bibinfo{person}{Zhenhe Chen}, {and} \bibinfo{person}{Jagath Samarabandu}.} \bibinfo{year}{2009}\natexlab{}.
\newblock \showarticletitle{Robust and efficient feature tracking for indoor navigation}.
\newblock \bibinfo{journal}{\emph{IEEE Transactions on Systems, Man, and Cybernetics, Part B (Cybernetics)}} \bibinfo{volume}{39}, \bibinfo{number}{3} (\bibinfo{year}{2009}), \bibinfo{pages}{658--671}.
\newblock


\bibitem[Saha et~al\mbox{.}(2019)]%
        {saha2019closing}
\bibfield{author}{\bibinfo{person}{Manaswi Saha}, \bibinfo{person}{Alexander~J Fiannaca}, \bibinfo{person}{Melanie Kneisel}, \bibinfo{person}{Edward Cutrell}, {and} \bibinfo{person}{Meredith~Ringel Morris}.} \bibinfo{year}{2019}\natexlab{}.
\newblock \showarticletitle{Closing the Gap: Designing for the Last-Few-Meters Wayfinding Problem for People with Visual Impairments}. In \bibinfo{booktitle}{\emph{The 21st International ACM SIGACCESS Conference on Computers and Accessibility}}. \bibinfo{pages}{222--235}.
\newblock


\bibitem[{\v{S}}akaja(2020)]%
        {vsakaja2020non}
\bibfield{author}{\bibinfo{person}{Laura {\v{S}}akaja}.} \bibinfo{year}{2020}\natexlab{}.
\newblock \showarticletitle{The non-visual image of the city: How blind and visually impaired white cane users conceptualize urban space}.
\newblock \bibinfo{journal}{\emph{Social \& cultural geography}} \bibinfo{volume}{21}, \bibinfo{number}{6} (\bibinfo{year}{2020}), \bibinfo{pages}{862--886}.
\newblock


\bibitem[Sato et~al\mbox{.}(2017)]%
        {sato2017navcog3}
\bibfield{author}{\bibinfo{person}{Daisuke Sato}, \bibinfo{person}{Uran Oh}, \bibinfo{person}{Kakuya Naito}, \bibinfo{person}{Hironobu Takagi}, \bibinfo{person}{Kris Kitani}, {and} \bibinfo{person}{Chieko Asakawa}.} \bibinfo{year}{2017}\natexlab{}.
\newblock \showarticletitle{NavCog3: An evaluation of a smartphone-based blind indoor navigation assistant with semantic features in a large-scale environment}. In \bibinfo{booktitle}{\emph{Proceedings of the 19th International ACM SIGACCESS Conference on Computers and Accessibility}}. \bibinfo{pages}{270--279}.
\newblock


\bibitem[Tekin and Coughlan(2010)]%
        {Tekin2010Barcode}
\bibfield{author}{\bibinfo{person}{Ender Tekin} {and} \bibinfo{person}{James~M. Coughlan}.} \bibinfo{year}{2010}\natexlab{}.
\newblock \showarticletitle{A mobile phone application enabling visually impaired users to find and read product barcodes}. In \bibinfo{booktitle}{\emph{Computers Helping People with Special Needs}}, \bibfield{editor}{\bibinfo{person}{Klaus Miesenberger}, \bibinfo{person}{Joachim Klaus}, \bibinfo{person}{Wolfgang Zagler}, {and} \bibinfo{person}{Arthur Karshmer}} (Eds.). \bibinfo{publisher}{Springer Berlin Heidelberg}, \bibinfo{address}{Berlin, Heidelberg}, \bibinfo{pages}{290--295}.
\newblock
\showISBNx{978-3-642-14100-3}


\bibitem[Tversky(1993)]%
        {cognitive_colalge_1993}
\bibfield{author}{\bibinfo{person}{Barbara Tversky}.} \bibinfo{year}{1993}\natexlab{}.
\newblock \showarticletitle{Cognitive maps, cognitive collages, and spatial mental models}. In \bibinfo{booktitle}{\emph{Spatial Information Theory A Theoretical Basis for GIS}}, \bibfield{editor}{\bibinfo{person}{Andrew~U. Frank} {and} \bibinfo{person}{Irene Campari}} (Eds.). \bibinfo{publisher}{Springer Berlin Heidelberg}, \bibinfo{address}{Berlin, Heidelberg}, \bibinfo{pages}{14--24}.
\newblock
\showISBNx{978-3-540-47966-6}


\bibitem[Voulodimos et~al\mbox{.}(2018)]%
        {voulodimos2018deep}
\bibfield{author}{\bibinfo{person}{Athanasios Voulodimos}, \bibinfo{person}{Nikolaos Doulamis}, \bibinfo{person}{Anastasios Doulamis}, {and} \bibinfo{person}{Eftychios Protopapadakis}.} \bibinfo{year}{2018}\natexlab{}.
\newblock \showarticletitle{Deep learning for computer vision: A brief review}.
\newblock \bibinfo{journal}{\emph{Computational intelligence and neuroscience}} \bibinfo{volume}{2018}, \bibinfo{number}{1} (\bibinfo{year}{2018}), \bibinfo{pages}{7068349}.
\newblock


\bibitem[Wiggett-Barnard and Steel(2008)]%
        {wiggett2008experience}
\bibfield{author}{\bibinfo{person}{Cindy Wiggett-Barnard} {and} \bibinfo{person}{Henry Steel}.} \bibinfo{year}{2008}\natexlab{}.
\newblock \showarticletitle{The experience of owning a guide dog}.
\newblock \bibinfo{journal}{\emph{Disability and Rehabilitation}} \bibinfo{volume}{30}, \bibinfo{number}{14} (\bibinfo{year}{2008}), \bibinfo{pages}{1014--1026}.
\newblock


\bibitem[Xie et~al\mbox{.}(2023)]%
        {10.1145/3563657.3596019}
\bibfield{author}{\bibinfo{person}{Jingyi Xie}, \bibinfo{person}{Rui Yu}, \bibinfo{person}{Kaiming Cui}, \bibinfo{person}{Sooyeon Lee}, \bibinfo{person}{John~M. Carroll}, {and} \bibinfo{person}{Syed~Masum Billah}.} \bibinfo{year}{2023}\natexlab{}.
\newblock \showarticletitle{Are Two Heads Better than One? Investigating Remote Sighted Assistance with Paired Volunteers}. In \bibinfo{booktitle}{\emph{Proceedings of the 2023 ACM Designing Interactive Systems Conference}} (Pittsburgh, PA, USA) \emph{(\bibinfo{series}{DIS '23})}. \bibinfo{publisher}{Association for Computing Machinery}, \bibinfo{address}{New York, NY, USA}, \bibinfo{pages}{1810–1825}.
\newblock
\showISBNx{9781450398930}
\urldef\tempurl%
\url{https://doi.org/10.1145/3563657.3596019}
\showDOI{\tempurl}


\bibitem[Xie et~al\mbox{.}(2024a)]%
        {10.1145/3613904.3642030}
\bibfield{author}{\bibinfo{person}{Jingyi Xie}, \bibinfo{person}{Rui Yu}, \bibinfo{person}{He Zhang}, \bibinfo{person}{Sooyeon Lee}, \bibinfo{person}{Syed~Masum Billah}, {and} \bibinfo{person}{John~M. Carroll}.} \bibinfo{year}{2024}\natexlab{a}.
\newblock \showarticletitle{BubbleCam: Engaging Privacy in Remote Sighted Assistance}. In \bibinfo{booktitle}{\emph{Proceedings of the 2024 CHI Conference on Human Factors in Computing Systems}} (Honolulu, HI, USA) \emph{(\bibinfo{series}{CHI '24})}. \bibinfo{publisher}{Association for Computing Machinery}, \bibinfo{address}{New York, NY, USA}, Article \bibinfo{articleno}{48}, \bibinfo{numpages}{16}~pages.
\newblock
\showISBNx{9798400703300}
\urldef\tempurl%
\url{https://doi.org/10.1145/3613904.3642030}
\showDOI{\tempurl}


\bibitem[Xie et~al\mbox{.}(2024b)]%
        {xie2024emergingpracticeslargemultimodal}
\bibfield{author}{\bibinfo{person}{Jingyi Xie}, \bibinfo{person}{Rui Yu}, \bibinfo{person}{He Zhang}, \bibinfo{person}{Sooyeon Lee}, \bibinfo{person}{Syed~Masum Billah}, {and} \bibinfo{person}{John~M. Carroll}.} \bibinfo{year}{2024}\natexlab{b}.
\newblock \bibinfo{title}{Emerging Practices for Large Multimodal Model (LMM) Assistance for People with Visual Impairments: Implications for Design}.
\newblock
\newblock
\showeprint[arxiv]{2407.08882}~[cs.HC]
\urldef\tempurl%
\url{https://arxiv.org/abs/2407.08882}
\showURL{%
\tempurl}


\bibitem[Yu et~al\mbox{.}(2024)]%
        {yu2024human}
\bibfield{author}{\bibinfo{person}{Rui Yu}, \bibinfo{person}{Sooyeon Lee}, \bibinfo{person}{Jingyi Xie}, \bibinfo{person}{Syed~Masum Billah}, {and} \bibinfo{person}{John~M Carroll}.} \bibinfo{year}{2024}\natexlab{}.
\newblock \showarticletitle{Human--AI Collaboration for Remote Sighted Assistance: Perspectives from the LLM Era}.
\newblock \bibinfo{journal}{\emph{Future Internet}} \bibinfo{volume}{16}, \bibinfo{number}{7} (\bibinfo{year}{2024}), \bibinfo{pages}{254}.
\newblock
\urldef\tempurl%
\url{https://doi.org/10.3390/fi16070254}
\showDOI{\tempurl}


\bibitem[Yuan et~al\mbox{.}(2017)]%
        {10.1145/3134753}
\bibfield{author}{\bibinfo{person}{Chien~Wen Yuan}, \bibinfo{person}{Benjamin~V. Hanrahan}, \bibinfo{person}{Sooyeon Lee}, \bibinfo{person}{Mary~Beth Rosson}, {and} \bibinfo{person}{John~M. Carroll}.} \bibinfo{year}{2017}\natexlab{}.
\newblock \showarticletitle{I Didn't Know that You Knew I Knew: Collaborative Shopping Practices between People with Visual Impairment and People with Vision}.
\newblock \bibinfo{journal}{\emph{Proc. ACM Hum.-Comput. Interact.}} \bibinfo{volume}{1}, \bibinfo{number}{CSCW}, Article \bibinfo{articleno}{118} (\bibinfo{date}{Dec.} \bibinfo{year}{2017}), \bibinfo{numpages}{18}~pages.
\newblock
\urldef\tempurl%
\url{https://doi.org/10.1145/3134753}
\showDOI{\tempurl}


\bibitem[Zhang et~al\mbox{.}(2024a)]%
        {10494076}
\bibfield{author}{\bibinfo{person}{He Zhang}, \bibinfo{person}{Xinyang Li}, \bibinfo{person}{Yuanxi Sun}, \bibinfo{person}{Xinyi Fu}, \bibinfo{person}{Christine Qiu}, {and} \bibinfo{person}{John~M. Carroll}.} \bibinfo{year}{2024}\natexlab{a}.
\newblock \showarticletitle{VRMN-bD: A Multi-modal Natural Behavior Dataset of Immersive Human Fear Responses in VR Stand-up Interactive Games}. In \bibinfo{booktitle}{\emph{2024 IEEE Conference Virtual Reality and 3D User Interfaces (VR)}}. \bibinfo{pages}{320--330}.
\newblock
\urldef\tempurl%
\url{https://doi.org/10.1109/VR58804.2024.00054}
\showDOI{\tempurl}


\bibitem[Zhang et~al\mbox{.}(2024b)]%
        {zhang2024redefiningqualitativeanalysisai}
\bibfield{author}{\bibinfo{person}{He Zhang}, \bibinfo{person}{Chuhao Wu}, \bibinfo{person}{Jingyi Xie}, \bibinfo{person}{Yao Lyu}, \bibinfo{person}{Jie Cai}, {and} \bibinfo{person}{John~M. Carroll}.} \bibinfo{year}{2024}\natexlab{b}.
\newblock \bibinfo{title}{Redefining Qualitative Analysis in the AI Era: Utilizing ChatGPT for Efficient Thematic Analysis}.
\newblock
\newblock
\showeprint[arxiv]{2309.10771}~[cs.HC]
\urldef\tempurl%
\url{https://arxiv.org/abs/2309.10771}
\showURL{%
\tempurl}


\bibitem[Zientara et~al\mbox{.}(2017)]%
        {Zientara2017thirdEye}
\bibfield{author}{\bibinfo{person}{P.~A. Zientara}, \bibinfo{person}{S. Lee}, \bibinfo{person}{G.~H. Smith}, \bibinfo{person}{R. Brenner}, \bibinfo{person}{L. Itti}, \bibinfo{person}{M.~B. Rosson}, \bibinfo{person}{J.~M. Carroll}, \bibinfo{person}{K.~M. Irick}, {and} \bibinfo{person}{V. Narayanan}.} \bibinfo{year}{2017}\natexlab{}.
\newblock \showarticletitle{Third Eye: A shopping assistant for the visually impaired}.
\newblock \bibinfo{journal}{\emph{Computer}} \bibinfo{volume}{50}, \bibinfo{number}{02} (\bibinfo{date}{feb} \bibinfo{year}{2017}), \bibinfo{pages}{16--24}.
\newblock
\showISSN{1558-0814}
\urldef\tempurl%
\url{https://doi.org/10.1109/MC.2017.36}
\showDOI{\tempurl}


\end{thebibliography}

\appendix

\end{document}